\documentclass[11pt]{article} 
\usepackage{epsfig,amssymb}
\usepackage{amsmath}
\usepackage{natbib}
\usepackage{graphics}
\usepackage{graphicx}
\usepackage{dcolumn}
\usepackage{bm}
\usepackage{epsfig}

\begin{document}
\begin{center}

\Large{\bf The Pioneer Anomaly and Its Implications}

\vskip 15pt

\normalsize
\bigskip 

Slava G. Turyshev,$^a$
Michael Martin Nieto,$^b$
and 
John D. Anderson$^a$

\normalsize
\vskip 10pt

{\it $^a$Jet Propulsion Laboratory, 
California Institute of Technology, \\
4800 Oak Grove Drive, Pasadena, CA 91109, USA\\
$^b$Theoretical Division (MS-B285), 
Los Alamos National Laboratory,\\
University of California, Los Alamos, NM 87545, USA}

\vskip 20pt

\end{center}

\begin{abstract} 
The Pioneer 10/11 spacecraft yielded the most precise navigation in deep space to date.  However, their radio-metric tracking data has consistently indicated the presence of a small, anomalous, Doppler frequency drift. The drift is a blue-shift, uniformly changing with a rate of $\sim6\times 10^{-9}$~Hz/s and can be interpreted as a constant sunward acceleration of each particular spacecraft of $a_P  = (8.74 \pm 1.33)\times 10^{-10}$~m/s$^2$. The nature of this anomaly remains unexplained. 
Here we summarize our current knowledge of the discovered effect and review some of the mechanisms proposed for its explanation.  Currently we are preparing for the analysis of the entire set of the available Pioneer 10/11 Doppler data which may shed a new light on the origin of the anomaly.  We present a preliminary assessment of such an intriguing possibility. 
\end{abstract}
%
%

\section{The Pioneer Missions and the Anomaly}
\label{mission}

The Pioneer 10/11 missions, launched on 2 Mar 1972 (Pioneer
10) and 5 Apr 1973 (Pioneer 11), respectively, were the first spacecraft to explore the outer solar system \citep{pioprd}. Their objectives were to conduct, during the 1972-73 Jovian opportunities, exploratory investigation beyond the orbit of Mars of the interplanetary medium, the nature of the asteroid belt, and the environmental and atmospheric characteristics of Jupiter and Saturn (for Pioneer 11). After Jupiter and Saturn encounters, the craft followed escape hyperbolic orbits near the plane of the ecliptic to opposite sides of the solar system, continuing on their extended missions. Pioneer 10 eventually became the first man-made object to leave the solar system.  The last coherent telemetry data point was obtained from Pioneer 10 on 27 Apr 2002 when the craft was 80~AU from the Sun. (Pioneer 11 sent its last useful data on 1 Oct 1990 while at $\sim30$ AU from the Sun.) 

The Pioneers were excellent for precise celestial mechanics experiments \citep{stanford}.  However, by 1980, when Pioneer 10 had  already passed a distance of  $\sim$ 20 AU from the Sun and the acceleration contribution from solar-radiation pressure on the craft (away from the Sun) had decreased to less than $4\times 10^{-10}$ m/s$^2$, the anomalous acceleration (towards the Sun) became evident in the data.  The detailed study of this anomaly  \citep{pioprd} led to a better understanding of its properties, as summarized in the next section.

\subsection{Summary of the Pioneer Anomaly}
\label{anomaly}

The analysis of the Pioneer 10 and 11 data \citep{pioprl,pioprd} demonstrated the presence of an anomalous, Doppler frequency blue-shift drift, uniformly changing with a rate of 
$\dot{f}_P \sim 6\times 10^{-9}$~Hz/s \citep{stanford}.
To understand the phenomenology of the effect, consider $f_{\tt obs}$, the frequency of the re-transmitted signal observed by a DSN antenna, and $f_{\tt model}$,  the predicted frequency  of that signal. The observed, two-way (round-trip) anomalous effect can be expressed to first order in $v/c$ as  
$\left[f_{\tt obs}(t)- f_{\tt model}(t)\right]_{\tt DSN}
=  -2\dot{f}_P\,t,$ 
with  $f_{\tt model}$ being the modeled frequency change due to conventional forces influencing the spacecraft's motion.  

After accounting for the gravitational and other large forces included in standard orbit determination programs this translates to 
\begin{eqnarray}
\left[f_{\tt obs}(t)- f_{\tt model}(t)\right]_{\tt DSN}
= -f_{0}\frac{2a_P~t}{c}. 
\label{eq:delta_nu_syst}
\end{eqnarray}
Here $f_{0}$ is the reference frequency \citep{pioprd}. 
 
Furthermore, after accounting for all {\it known} (not modeled) sources of systematic error (discussed in \cite{pioprd,old4}), conclusion reached was that there exists an anomalous sunward constant acceleration signal of 
\begin{equation}
a_P=(8.74\pm1.33)\times 10^{-10}~~{\rm m/s}^2. 
\end{equation}

For the most detailed analysis of the Pioneer anomaly to date, \cite{pioprd} used the following Pioneer 10/11 Doppler data: 
\begin{itemize}
\item Pioneer 10: The data set had 20,055 data points obtained between 3 Jan 1987 and 22 Jul 1998 and covering heliocentric distances  $\sim40-70.5$~AU.   
\item Pioneer 11: The data set had 10,616 data points obtained between 5 Jan 1987 to 1 Oct 1990 and covering heliocentric distances $\sim22.42-31.7$~AU. 
\end{itemize}

By now several studies of the Pioneer Doppler navigational data have demonstrated that the anomaly is unambiguously present in the Pioneer 10 and 11 data. These studies were performed with three independent (and different!) navigational computer programs 
\citep{pioprl,pioprd,markwardt}.  Namely: 
{}
\begin{itemize}
\item the JPL's Orbit Determination Program (ODP) developed in 1980-2005,
\item The Aerospace Corporation's  CHASPM code extended for deep space navigation \citep{pioprl,pioprd}, and finally 
\item a code written in the Goddard Space Flight Center \citep{markwardt} that was used to analyze Pioneer 10 data for the period 1987-1994 obtained from the National Space Science Data Center ({\tt http://nssdc.gsfc.nasa.gov/}). 
\end{itemize}

The recent analyses of the Pioneer 10 and 11 radio-metric data \citep{pioprl,pioprd,markwardt,stanford} have established the following basic properties of the Pioneer anomaly:
{}
\begin{itemize}
\item {\it Direction:} Within the 10 dbm bandwidth of the high-gain antennae, $a_P$ behaves as a line-of-sight constant acceleration of the craft toward the Sun.
\item {\it Distance:} It is unclear how far out the anomaly goes, but the Pioneer 10 data supports its presence at distances up to $\sim$70~AU from the Sun. The Pioneer 11 data shows the presence of the anomaly as close in as $\sim$20 AU. 
\item {\it Constancy:}~Both temporal and spatial variations of the anomaly's magnitude are of order 10\% for each craft, while formal errors are significantly smaller.
\end{itemize}

This information was used as guidance in investigating the applicability of proposals to explain the Pioneer anomaly using both conventional and `new' physical mechanisms. In the next section we will briefly review these proposals.  

\section{Recent Efforts to Explain the Anomaly}
\label{sec:explain}
\subsection{Conventional physics mechanisms}

Efforts to explain the anomaly were originally focused on conventional physics mechanisms generated on-board, such as gas leaks from the propulsion system or a recoil force due to the on-board thermal power inventory. So far, these mechanisms have been found to be either not strong enough to explain the magnitude of the anomaly or else to exhibit significant temporal or spatial variations contradicting the known properties of the anomaly presented above \citep{pioprd,old4}.

A number of other conventional physics possibilities have also been addressed.  In particular, it has been proposed that Kuiper Belt Objects or dust could explain the anomaly by (i) a gravitational acceleration, (ii) an additional drag force (resistance) and (iii) a frequency shift of the radio signals proportional to the distance.

Of course, one of the most natural mechanism to generate a putative physical force is the gravitational attraction due to a known mass distribution in the outer solar system; for instance, due to Kuiper Belt Objects or dust. However, possible density distributions for the Kuiper belt were studied in \citep{pioprd} and found to be incompatible with the discovered properties of the anomaly. Even worse, these distributions cannot circumvent the constraint from the undisturbed orbits of Mars and Jupiter \citep{pioprd}. The density of dust is not large enough to produce a gravitational acceleration on the order of $a_P$ \citep{pioprd,mmn05} and also it varies greatly within the Kuiper belt, precluding any constant acceleration. Hence, a gravitational attraction by the Kuiper belt can, to a large extent, be ruled out.

Also, the data from the inner parts of the solar system taken by the Pioneer 10/11 dust detectors favors a spherical distribution of dust over a disk in this inner region. Ulysses and Galileo measurements in the inner solar system find very few dust grains in the $10^{-18}-10^{-12}$~kg range \citep{dust}. IR observations rule out more than 0.3 Earth mass from Kuiper Belt dust in the trans-Neptunian region. The resistance caused by the interplanetary dust is too small to provide
support for the anomaly \citep{dust}.  Any dust-induced frequency shift of the carrier signal is also ruled out.

Finally we note that, motivated by the numerical coincidence $a_P\simeq cH_0$, where $c$ the speed of light and $H_0$ is the Hubble constant at the present time, there have been many attempts to explain the anomaly in terms of the expansion of the Universe.  \cite{pioprd}  shown that such a mechanism would produce an opposite sign for the effect. A study of the effect of cosmic acceleration on the radio signals rather than on the spacecraft themselves was also undertaken.  This mechanism might be able to overcome the apparent conflict that $a_P$ presents to modern solar system planetary ephemerides \citep{pioprl,pioprd}.

\subsection{Possibility for new physics?} 
\label{sec:new-ph}

The apparent difficulty to explain the anomaly within standard physics became a motivation to look for `new physics.' So far these attempts have not produced a clearly viable mechanism for the anomaly.  In particular:

The physics of MOND represents an interesting possibility with its phenomenological long-range modification of gravity invoked to explain the rotation curves of galaxies \citep{Milgrom,Bekenstein}. However, the numerical value for the MONDian acceleration $a_0$ is almost on order of magnitude smaller then $a_P$ and is not likely to be observed on the scales of the solar system.  

 There is also an attempt to explain the anomaly in the framework of a non–symmetric gravitational theory \citep{moffat}. It has been argued that a scalar field with a suitable potential could account for $a_P$ \citep{moffat05}. A modification of the gravitational field equations for a metric gravity field, by introducing a general linear relation between the Einstein tensor and the energy-momentum tensor has also been claimed to account for $a_P$ \citep{serge2}. 

Various distributions of dark matter in the Solar system have been proposed to explain the anomaly, e.g., dark matter distributed in a form of a disk in the outer solar system with of a density of $\sim4\times10^{-16}$~kg/m$^3$, yielding the wanted effect. Dark matter in the form of `mirror matter' \citep{FootVolkas01}  is one example. However, it would have to be a special smooth distribution of dark matter that is not gravitationally modulated as normal matter so obviously is.

\cite{orfeu04} have shown that a generic scalar field cannot explain $a_P$; on the other hand they proposed that a non-uniformly-coupled scalar might produce the wanted effect. Although brane-world models with large extra dimensions offer a richer phenomenology than standard scalar-tensor theories, it is difficult to find a convincing explanation for the Pioneer anomaly \citep{orfeu_jorge05}. Other ideas include Yuka\-wa-like or higher order corrections to the Newtonian potential, a theory of conformal gravity with dynamical mass generation, including the Higgs scalar \citep{pioprd}, and others. 

Concluding, we must say that there are many interesting ideas, however, most of them need more work before they can be considered to be viable.

\subsection{Search for independent confirmation}
\label{sec:experiemnts}

Attempts to verify the anomaly using other spacecraft proved disappointing. This is because the Voyager, Galileo, Ulysses, and Cassini  spacecraft navigation data all have their own individual difficulties for use in an independent test of the anomaly. In addition, many of the deep space missions that are currently being considered either may not provide the needed navigational accuracy and trajectory stability sensitive to accelerations of under $10^{-10}$~m/s$^2$ or else they will have significant on-board systematics that mask the anomaly. A requirement to have an escape hyperbolic trajectory makes a number of other missions \citep{stanford} less able to directly test $a_P$.  Although these missions all have excellent scientific goals and technologies, nevertheless, their orbits lend them a less advantageous position to conduct a precise test of the detected anomaly. 

A number of alternative ground-based verifications of the anomaly have also been considered; for example, using Very Long Baseline Interferometry (VLBI) astrometric observations.  However, the trajectories of spacecraft like the Pioneers, with small proper motions in the sky, make it presently impossible to use VLBI in accurately isolating an anomalous sunward acceleration of the size of $a_P$.

To summarize, the origin of the Pioneer anomaly remains unclear. 


\section{Analysis of the Entire Set of Existing Pioneer 10/11 Data}
\label{edata}

\cite{stanford} advocated that analysis of the entire set of existing Pioneer data, obtained from launch to the last useful data received from Pioneer 10 in April 2002.  This data could yield critical new information about the anomaly \citep{stanford,edata-mmn05}.  An effort to retrieve the early data, existing on various obsolete-format media, and transfer it to modern DVDs is being initiated at JPL; as of October 2005 this transfer is nearing completion.  

This transfer will enable us to investigate the Pioneer anomaly with the entire available Pioneer 10/11 radio-metric Doppler data.  The objectives of this new investigation of the Pioneer anomaly will be three-fold: (i) to study the physics of the planetary encounters, (ii) to analyze the early mission data, and (iii) to study the temporal evolution of $a_P$ with the entire data set.

\subsection{ Planetary encounters} 

\begin{figure*}[t!]
 \begin{center}
\noindent   
\psfig{figure=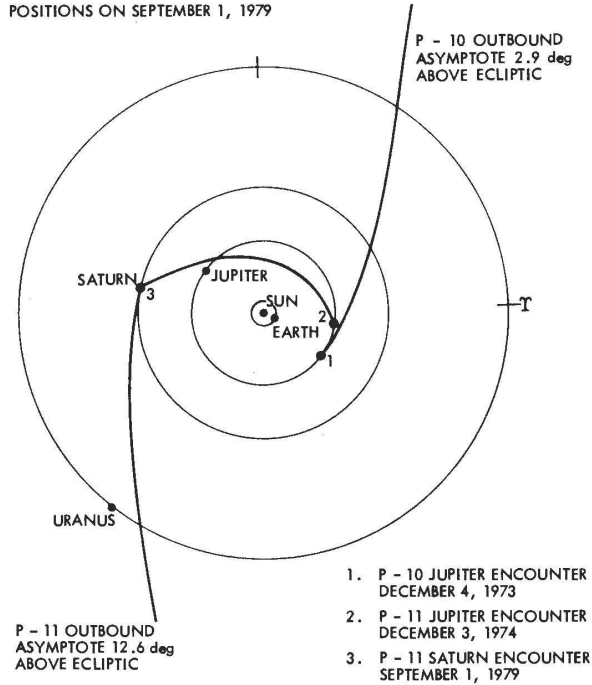,width=8.0cm}
\end{center}
\vskip -10pt 
  \caption{Heliocentric geometry of Pioneer 10 and 11 trajectories.
 \label{fig:pio-inner-saturn}}
\vskip -10pt 
\end{figure*}

The early Pioneer 10 and 11 data (before 1987) were never analyzed in detail, especially with a regard for systematics.  However, for Pioneer 10 an approximately constant anomalous acceleration seems to exist in the data as close in as 27 AU from the Sun.  Pioneer 11, beginning just after Jupiter flyby, finds a small value for the anomaly during the Jupiter-Saturn cruise phase in the interior of the solar.  But right at Saturn encounter (see Figure~\ref{fig:pio-inner-saturn}), when the craft passed into an hyperbolic escape orbit,  there was a fast increase in the anomaly where-after it settled into the canonical value. (See \citep{edata-mmn05} for more detail).

%

We first plan to study the Saturn encounter for Pioneer 11. We plan to use the data for approximately two years surrounding this event. If successful, we should be able to find more information on the mechanism that led to the onset of the anomaly during the flyby. The Jovian encounters are also of significant interest. However, they were in the region much too close to the Sun. Thus one expects large contribution from the standard sources of acceleration  noise that exist at the heliocentric distances $\sim5$~AU. Nevertheless we plan to use a similar strategy as with the Saturn encounter and will attempt to make full use of the data available.

\subsection{The earlier data} 

We plan to analyze the early parts of the trajectories of the Pioneers with the goal of determining the true direction of the Pioneer anomaly and possibly its origin.  While in deep space, for standard antennae without good 3-D navigation, there are four possible distinct directions \citep{stanford}, namely: (i) the sunward direction that would indicate a force originating from the Sun, (ii) the Earth-pointing direction that would be due to an anomaly in the frequency standards, (iii) the direction alone the velocity vector that would indicate an inertial or drag force, or (iv) the spin-axis direction that would indicate an on-board systematic.  Therefore, analysis of the earlier data would be critical in helping to establish a precise 3-dimensional, time history of the effect. 

\begin{figure*}[t!]
 \begin{center}
\noindent   
\psfig{figure=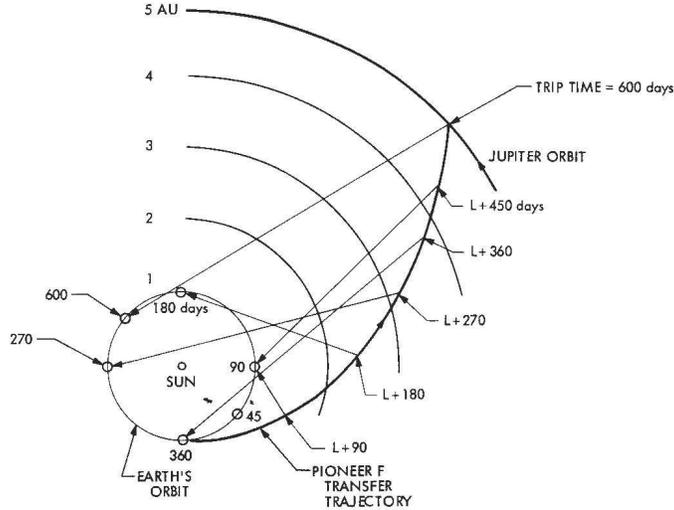,width=9.0cm}
\end{center}
\vskip -10pt 
  \caption{
Proposed directions (along the spin and antenna axes) from the Pioneer F spacecraft (to become Pioneer 10) toward the Earth.
 \label{fig:antenna-point}}
\vskip -10pt 
\end{figure*} 
%

During the flight in the inner solar system, an Earth-pointing attitude was necessary to enable the narrow beam of the highgain antenna of the spacecraft to illuminate the Earth and to maintain effective communications. Figure~\ref{fig:antenna-point} depicts the directions along the spacecraft's spin and antenna axes toward the Earth. The geocentric longitude of the craft varied continuously throughout the mission and, therefore, it was necessary to make numerous attitude adjustments. Since the spacecraft high-gain antenna has a half-power beamwidth of $\sim3.5^\circ$, many spin-axis orientation maneuvers were necessary to compensate for both the relative movement of the craft relative to the Earth, and also for the precession caused by solar pressure, which was 0.2$^\circ$ per day during the early part of the mission. In addition, to provide the planned encounter trajectories, some adjustments of the velocity vector were performed during the interplanetary flight by generating thrust in a particular direction.  Therefore, with thrusters in a fixed relationship to the spacecraft, reorientations of the spacecraft were also necessary.

\subsection{Entire data span} 

The same investigation could study the temporal evolution of the magnitude of the anomaly.  Thus, if the anomaly is due to the on-board nuclear fuel inventory (Pu-238) available on the vehicles and related heat recoil force, one expects that a decrease in the anomaly's magnitude will be correlated with Pu decay with half-life of 87.74 years.  The analyses of 11.5 years of data \citep{pioprd,old4} found no support for a thermal mechanism.   However, the available 30-year interval of data might be able demonstrate the effect of a $\sim24$\% reduction in the heat contribution to the craft's acceleration.  
We plan to analyze the entire set of available data in an attempt to determine whether or not the anomaly is due to its on-board nuclear fuel inventory and related heat radiation or to an other mechanism.

\section{Conclusion}
\label{conclude}

The existence of the Pioneer anomaly is no longer in doubt.  Further, after much understandable hesitancy, a steadily growing part of the community has concluded that the anomaly should be subject to interpretation.  The results of the investigation of the Pioneer anomaly would be win-win; improved navigational protocols for deep space at the least, exciting new physics at the best.  Finally, a strong international collaboration would be an additional outcome of the proposed program of understanding the Pioneer anomaly.

\subsection*{Acknowledgements}

The work of SGT and JDA  was carried out at the Jet Propulsion Laboratory, California Institute of Technology, under a contract with the National Aeronautics and Space Administration.  MMN acknowledges support by the U.S. Department of Energy.




\begin{thebibliography}{99}    

{\small

\bibitem[\protect\astroncite{Anderson et~al.}{1998}]{pioprl}  
Anderson, J.\,D. et~al. 1998, 
Phys. Rev. Lett. 81, 2858, gr-qc/9808081

\bibitem[\protect\astroncite{Anderson et~al.}{2002a}]{pioprd} 
Anderson, J.\,D. et~al. 2002a,
Phys. Rev. D. 65, 082004/1-50, gr-qc/0104064

\bibitem[\protect\astroncite{Anderson et~al.}{2002b}]{old4} 
Anderson, J.\,D. et~al. 2002b, 
Mod. Phys. Lett. A17, 875, gr-qc/0107022

\bibitem[\protect\astroncite{Bekenstein}{2004}]{Bekenstein}
Bekenstein,~J.\,D. 2004, 
astro-ph/0403694

\bibitem[\protect\astroncite{Bertolami \& Paramos}{2004}]{orfeu04} 
Bertolami,~O., P\'aramos,~J. 2004,
Class. Quant. Grav. 21, 3309,  gr-qc/0310101

\bibitem[\protect\astroncite{Bertolami \& Paramos}{2005}]{orfeu_jorge05} 
Bertolami,~O., P\'aramos,~J. 2005,
Phys. Rev. D 71, 023521, astro-ph/0408216

\bibitem[\protect\astroncite{Foot \& Volkas}{2001}]{FootVolkas01} 
Foot,~R., Volkas,~R.\,R. 2001, 
Phys. Lett. B 517, 13, hep-ph/0108051

\bibitem[\protect\astroncite{Jaekel \& Reynaud}{2005}]{serge2} 
Jaekel,~M.-T., Reynaud,~S. 2005,
Class.\,Quant.\,Grav.\,22, 2135, gr-qc/0502007

\bibitem[\protect\astroncite{Markwardt}{2002}]{markwardt}
Markwardt,~C.
2002,  gr-qc/0208046 

\bibitem[\protect\astroncite{Milgrom}{2004}]{Milgrom}
Milgrom,~M. 2001, 
Acta Phys. Pol. B 32, 3613

\bibitem[\protect\astroncite{Moffat}{2004}]{moffat}
Moffat,~J.\,W. 2004,
gr-qc:0405076

\bibitem[\protect\astroncite{Moffat}{2005}]{moffat05} 
Moffat,~J.\,W. 2005,
gr-qc/0506021

\bibitem[\protect\astroncite{Nieto et~al.}{2005}]{dust} 
Nieto,~M.\,M. et~al. 2005,
Phys. Lett. B 613, 11, astro-ph/0501626

\bibitem[\protect\astroncite{Nieto}{2005}]{mmn05} 
Nieto,~M.\,M. 2005,
astro-ph/0506281

\bibitem[\protect\astroncite{Nieto \& Anderson}{2005}]{edata-mmn05} 
Nieto,~M.\,M., Anderson,~J.\,D. 2005,
gr-qc/0507052

\bibitem[\protect\astroncite{Turyshev et~al.}{2005a}]{stanford} 
Turyshev,~S.\,G. et~al. 2005a,  
Stanford e-Conf \#C041213, \#0310, 
gr-qc/0503021

\bibitem[\protect\astroncite{Turyshev et~al.}{2005b}]{problem_set_05} Turyshev,~S.\,G. et~al. 2005b,  
Amer. J. Phys., to be published, physics/0502123
}
\end{thebibliography}



\end{document}